\documentclass[aps,prd,onecolumn,groupedaddress,showpacs,nofootinbib,amssymb]{revtex4}
\usepackage[dvips]{graphicx}
\usepackage{amssymb}
\usepackage{diagbox, float}
\usepackage{amsmath}
\usepackage{graphicx,color}
\usepackage{amsfonts}
\usepackage{bm}
\usepackage{cancel}
\usepackage{comment}

\newcommand\be{\begin{equation}}
\newcommand\ee{\end{equation}}

\newcommand{\bea}{\begin{eqnarray}}
\newcommand{\eea}{\end{eqnarray}}

\allowdisplaybreaks[4]

\begin{document}

\title{Non-flat Universe with Tsallis holographic dark energy}
\author{Artyom V. Astashenok}
\email{aastashenok@kantiana.ru}
\author{Alexander S. Tepliakov}
\affiliation{I. Kant Baltic Federal University\\
236041, Kaliningrad, Russia\\}%

\tolerance=5000

\begin{abstract}

The cosmological evolution of non-flat universe filled with general Tsallis holographic dark energy is considered. We assumed that the density of the holographic component is defined as $\sim L^{2\gamma - 4}$ where $\gamma$ is the nonadditivity parameter and $L$ is the length of the event horizon. The influence of possible interaction between dark matter and holographic dark energy is investigated. We also considered asymptotic evolution of universe from an early matter dominated epoch and established the existence of de Sitter attractor for $\gamma<1$.

\end{abstract}


\maketitle

\section{Introduction}

One of the key challenges of modern physics remains the explanation of the accelerated expansion of the Universe, recorded in observations of type Ia supernovae ~\cite{1,2}. Observational data demonstrate a discrepancy between the observed dependence of the luminosity of these objects on the redshift and the theoretical predictions of the standard cosmological model, which takes into account only baryonic matter and radiation. To reconcile the theory with experiment, it is necessary to introduce an additional component - dark energy, characterized by a high degree of spatial homogeneity and negative pressure. According to modern astrophysical data \cite{Amanullah}, \cite{Blake}, \cite{LCDM-1}, \cite{LCDM-2}, \cite{LCDM-3}, \cite{LCDM-4}, \cite{LCDM-5}, \cite{LCDM-6}, \cite{LCDM-7}, the most successful model describing the dynamics of the Universe is the $\Lambda$CDM model, where dark energy is represented by the cosmological constant $\Lambda$. However, despite its empirical success, this model faces serious conceptual difficulties. First of all, this is the problem of the smallness of the cosmological constant: theoretical estimates of the vacuum energy within the framework of quantum field theory exceed the observed value by 120 orders of magnitude. An additional complication is the so-called coincidence problem: the current values of the densities of dark energy and matter turn out to be of the same order of magnitude, which requires an explanation of the mechanism of their temporal correlation.

The problem of dark energy has stimulated the development of alternative approaches to its explanation. One of the promising directions is the quintes\-se\-nce theory, within which dark energy is interpreted as a scalar field with an effective state parameter $-1<w<-1/3$ \cite{Caldwell}, \cite{Steinhardt}, \cite{Ferramacho}, \cite{Caldwell-2}. An important feature of quintessential models is the time dependence of the field energy density, which allows constructing so-called "tracking" solutions. These solu\-ti\-ons provide dynamic coordination of the densities of dark energy and matter at different stages of the evolution of the Universe, which potentially solves the coincidence problem.

Significant progress has also been made in the framework of modified gravity theories \cite{Capozziello}, \cite{Odintsov}, \cite{Turner}. In such models, the accelerated expansion of the Universe is explained by additional terms in the gravitational action that go beyond the standard description through the curvature of space-time. This approach provides a rich mathematical apparatus for the analysis of alternative mechanisms of cosmological dynamics.

The concept of holographic dark energy deserves special attention. It is discussed in detail in the review by \cite{Wang} and is based on the holographic principle \cite{3}, \cite{4}, \cite{5}. This principle states that the physical information inside a volume can be completely described by quantities specified on its boundary. In the context of cosmology, this allows expressing the density of dark energy through two fundamental quantities: the Planck mass $M_p$ and the characteristic scale $L$. The choice of the scale $L$ (e.g., the particle horizon, the event horizon, or the inverse of the Hubble parameter) plays a key role. Using the event horizon as a scale may seem paradoxical, since it implies dependence on future conditions, but such "teleological" boundary conditions are already used in black hole physics \cite{Novik}. Thus, this approach does not represent a logical contradiction, but rather demonstrates the nontrivial nature of cosmological processes. For the classical holographic model, the energy density is expressed as $\sim L^{-2}$ in Planck units.

A generalization of the classical model of holographic dark energy was proposed in \cite{Tsallis-2} (see also \cite{Tsallis}), where an approach based on the generalized Tsallis entropy is considered. This direction was further developed in the studies of \cite{Tavayef}, \cite{Jahromi}, \cite{Nojiri}, where the properties of the holographic Tsallis dark energy are studied in various cosmological contexts. We demonstrated the consistency of such a model, according to updated observational data \cite{Astashenok1} in this work the ranges of parameters of the Tsallis holographic dark energy model were determined and shown, for which the model corresponds to observational data at the level of the $\Lambda$CDM model. It looks objectively valid: every analysis reasonable in numerical applications requires the stability of the model to be considered. The classical way to demonstrate stability is to calculate the square of the speed of sound, however, due to the specificity of the definition of holographic dark energy, this method is unsuitable. In the work \cite{Astashenok2} we showed that it is important to take into account scalar perturbations also for the chosen characteristic scale at the boundary of the Universe, which leads to stable behavior of the model.

An important aspect is the possibility of interaction between matter and holographic dark energy, which was studied in \cite{Qihong}. Within the framework of such interaction ($\sim H\rho_{de}$), dark energy can decay, which leads to a redist\-ri\-bu\-tion of energy density between the components. At a certain intensity of interaction, the shares of matter and dark energy density stabilize, forming a stable balance. This dynamic equilibrium contributes to the transition of the Universe to the quasi-de Sitter expansion regime, thereby preventing possible singularities in its future evolution. In \cite{Astashenok3}, we came to the conclusion that taking into account the interaction can lead to phantomization of dark energy without the emergence of singularities in the future, which distinguishes this model from scenarios with phantom energy with a constant state parameter $w<-1$.

An additional way to analyze dark energy models is dynamic analysis, which allows us to determine the critical points of the system of cosmological equations and the nature of the behavior of the Universe near these points. In \cite{Astashenok4} we determined critical points in the model of generalized holographic dark energy with possible interaction with matter. It was shown that the fraction of dark energy can tend to 1 or to a constant value less than 1 due to the establishment of a dynamic equilibrium between matter and dark energy. This opens up prospects for explaining the coincidence problem in cosmology.

Finally, the contribution of spatial curvature to the Friedmann equations cannot be ruled out, although observational data limit its value to a value significantly smaller than other energy components. The study of universes with minimal spatial curvature, consistent with inflationary models and astro\-phy\-sical observations, remains a pressing task. Such studies are not only of academic value, but also contribute to a deeper understanding of the fundamental properties of the Universe.

In this work, we analyze  the cosmological evolution of the non-flat universe filled by Tsallis holographic dark energy with various values of parameters. Firstly we briefly invesigate the case when no interaction between matter and holographic dark energy. Our consideration shows that universe can approach the quasi-de Sitter regime for some parameters. Another scenarios include expansion of the universe according to power law with Hubble parameter $H\rightarrow 0$ and future singularuty when scale factor blows out. Then we consider the case of interaction between dark energy and matter. Finally we analyse the asymptotical evolution of the universe from early era of matter domination. For $\gamma<1$ the asymtotic de Sitter regime take place in a case of without interaction between matter and holographic dark energy. As showed interaction can change this simple picture.   

\section{Basic equations}

For Friedmann universe with arbitrary curvature, the metric can be written in the following form:
\begin{equation}
    ds^2 = dt^2 - \frac{a^2(t)}{1-kr^2}\left(dr^2 + r^2 d\Omega^2\right),
\end{equation}
where $a(t)$ is the scale factor and $k$ is the curvature of the spacetime, $k=0$ corresponds to spatially flat Universe, two other cases are $k=\pm 1$.

We use the natural system of units ($8\pi G = c = 1$) in which cosmological equations are 
\begin{equation}
     H^2 = \frac{\rho }{3} - \frac{k}{a^2},
\end{equation}
\begin{equation}
    2\dot{H} + 3H^2 = -p -\frac{k}{a^2}.
\end{equation}
Here $H$ is the Hubble parameter $H\equiv \frac{\dot{a}}{a}$, $\rho$ and $p$ are the total energy density and pressure of all components of the Universe.  For dark energy density, we take the following expression:
\begin{equation}
   \rho_{de} = \frac{3C^2}{L^{4-2\gamma}},
\end{equation}
where $L$ is some scale in the Universe and $\gamma$ is a parameter of non-additivity and $C$ is the constant. If $\gamma=1$ the $C$ is non-dimensional. For $L$ we choose the distance as the event horizon i.e.
\begin{equation}
   L = ar_{h}(t),
\end{equation}
where $r_h(t)$ is the comoving distance to the horizon which can be defined from the relation
\begin{equation}
   R_h = \int_{t}^{\infty } \frac{dt}{a} = \int_{0}^{r_{h}(t)}\frac{dr}{\sqrt{1 - kr^2}}. 
\end{equation}
The time derivative of $R_h(t)$ is 
\begin{equation}
   \dot{R}{_h} = -\frac{1}{a}.
\end{equation}
Then we can write that
\begin{equation}
   r_{h}(t) = \frac{1}{\sqrt{k}}\sin\theta,\quad \theta = \sqrt{k} R_h.
\end{equation}
For fraction of dark energy density we have
\begin{equation}
   \Omega_{de} = \frac{C^2}{L^{4-2\gamma}H^2}.
\end{equation}

To solve the aforementioned cosmological equations, one needs to derive the pressure of dark energy. We can use for this purpose the continuity equation
$$
\dot{\rho}_{de} + 3 H (\rho_{de} + p_{de}) = 0.
$$
The derivative of the dark energy density can be calculated using Eqs. (7), (8):
\begin{equation}
   \dot{\rho}_{de} = (2\gamma - 4)H\rho_{de}(1 - \frac{\sqrt{k}}{aH}\frac{1}{\tan\theta}).
\end{equation}
For dynamical analysis of cosmological evolution we introduce for convenience the dimensionless quantities
\begin{equation}
x = \frac{\rho_m}{3H^2},\: \:\:  y = \frac{\rho_{de}}{3H^2}, \: \:\: \Omega_k = \frac{k}{a^2H^2}.
\end{equation}
The $x$, $y$ and $\Omega_k$ satisfy the obvious equality
\begin{equation}
   x + y - \Omega_k = 1.
\end{equation}
We also add the dimensionless variable $u$ defined as
$u = L H$.
After simple technical calculations we can obtain the following system of equations for $x$, $y$ and $u$:
\begin{equation}
\frac{dx}{d \eta } = -xy(3+(2\gamma-4)(1 - \hat{\delta})),
\end{equation}
\begin{equation}
\frac{dy}{d \eta } = y(1-y)(3+(2\gamma-4)(1 - \hat{\delta})),
\end{equation}
\begin{equation}
\frac{du}{d\eta} = u\left(-\frac{x}{2} + y + (\gamma - 2) y (1-\hat{\delta})\right) - \cos{\theta}
\end{equation}
Here $\eta$ means the dimensionless time ($d\eta \equiv H dt$) and for brevity we introduced the function $\hat{\delta}$:
$$\hat{\delta} = \sqrt{x+y-1} \cot \theta$$
where
$$
\theta =  \arcsin\left [u \sqrt{x+y-1}\right].
$$
For flat universe $x+y=1$ and, therefore, $\theta=0$. Going to the limit $x+y\rightarrow 1$ for $\hat{\delta}$ we simply have $\hat{\delta} = 1/u$.  




\section{THDE model without interaction between \\ matter and dark energy}

We considered for illustration three values of $C$: 0.7, 1 and 1.2. For non-additivity parameter $\gamma$ the values $0.9$, $1$, $1.2$ and $1.5$ are taken. We investigate evolution of the Universe starting from current time, assuming that $\Omega_{de0}=0.73$. For $\Omega_k$ the value $0.01$ is taken. The Hubble parameter is measured in units of $\rho_0$ (therefore, $H_0 = 3^{-1/2}$ at current time). The results of our calculations for future and past are showed on fig. 1. It is convenient to use variable $z = a_{0}/a - 1$ instead of time. For past this variable has meaning of redshift.

\begin{figure}
    \centering
    \includegraphics[width=1\linewidth]{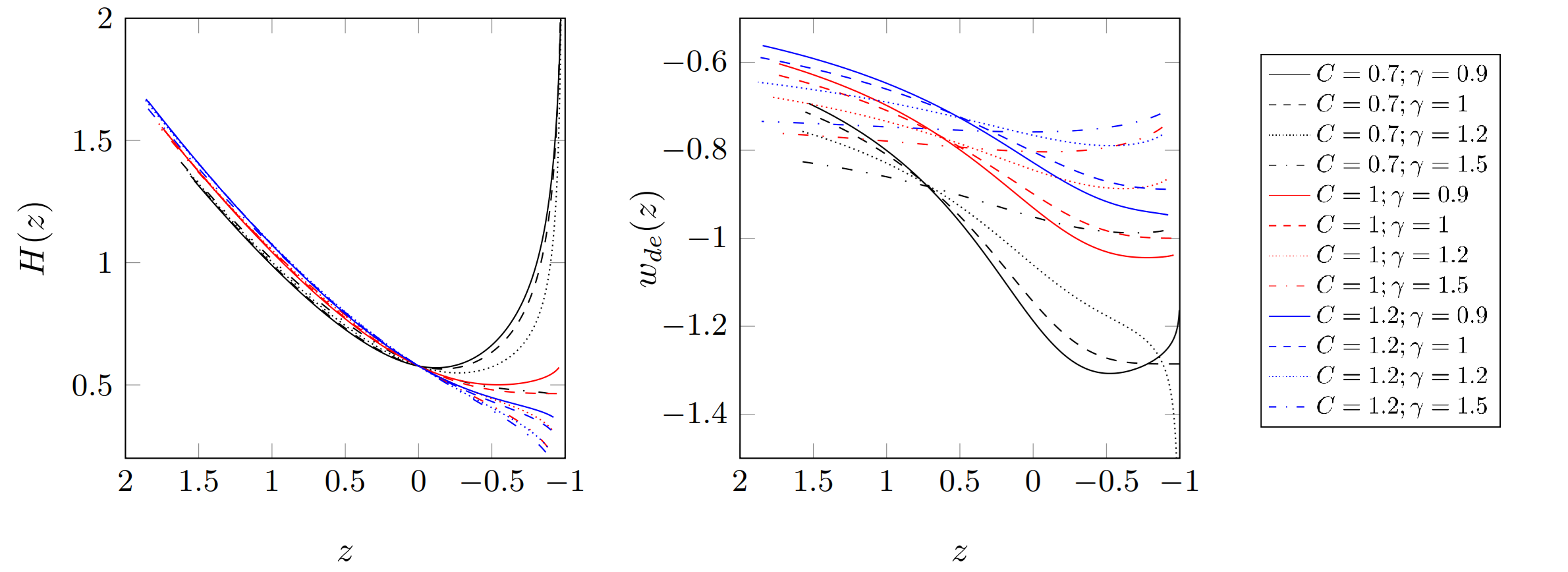}
    \caption{The dependence of Hubble parameter (in dimensionless units) from $z = a_0/a - 1$ (upper panel) and equation-of-state parameter $w_{de}=p_{de}/\rho_{de}$ for holographic dark energy (lower panel). Negative ``redshifts'' correspond to future times, $z\rightarrow 1$ corresponds to $a\rightarrow\infty$ or some finite moment of time (singularity). For $\gamma<1$, there is no singularity in future.}
    \label{fig1}
\end{figure}

We see that for $C=0.7$ and $0.9<\gamma<1.2$ the Hubble parameter rapidly increases. At first glance, this can be interpreted as a big rip singularity at a finite moment of time. Although, more detailed analysis shows that for $\gamma<1$ there is no singularity in future, and, for large times, $H\rightarrow\mbox{const}$. Additionally, at some moment in the past, the phantomization occurs. However,  for $\gamma=1.5$ $w_{de}>-1$ at every moment of time, and Universe expands with some finite rate at $t\rightarrow\infty$. For $C>1$, the situation is the same as for usual holographic dark energy with $\gamma=1$. The Hubble parameter decreases with time. Finally, in the case of $C=1$ we see phantomization in future for $\gamma=0.9$. For $\gamma>1$, the cosmological evolution is similar to the case of $C>1$. 

\section{THDE with interaction between \\ matter and dark energy}

We can propose that holographic dark energy and matter interact with each other. According to this assumption, we introduce some function $Q(\rho_m, \rho_{de}, H)$ in r.h.s. of continuity equation for each component:
\begin{equation}
\dot{\rho}_{m}+3H\rho_{m}=-Q,
\end{equation}

\begin{equation}
\dot{\rho}_{de}+3H(\rho_{de}+p_{de})=Q.
\end{equation}

In this case, the system of equations for dimensionless variables $x$, $y$ and $u$ takes the form:
\begin{equation}
\frac{dx}{d \eta } =q(1-x) -xy(3+(2\gamma-4)(1 - \hat{\delta})),
\end{equation}
\begin{equation}
\frac{dy}{d \eta } = y(1-y)(3+(2\gamma-4)(1 - \hat{\delta})) - qy,
\end{equation}
\begin{equation}
\frac{du}{d\eta} = u\left(-\frac{x}{2} + y + (\gamma - 2) y (1-\hat{\delta}) -\frac{q}{2}\right) - \cos{\theta},
\end{equation}
where the following dimensionless quantity is introduced:
\begin{equation}
   q = \frac{Q}{3H^3}
\end{equation}

Let's consider three types of interaction bwtween matter and dark energy.
\textbf{Model with $Q = H(\alpha \rho_m + \beta \rho_{de})$.} This choice corresponds to 
\begin{equation}
q = \alpha x + \beta y.
\end{equation}
We consider such $\alpha$ and $\beta$ which don't lead to unphysical behaviour of energy density for instance $\rho_{de}$ or $\rho_m$ become negative. 
\begin{figure}
    \centering
    \includegraphics[width=1\linewidth]{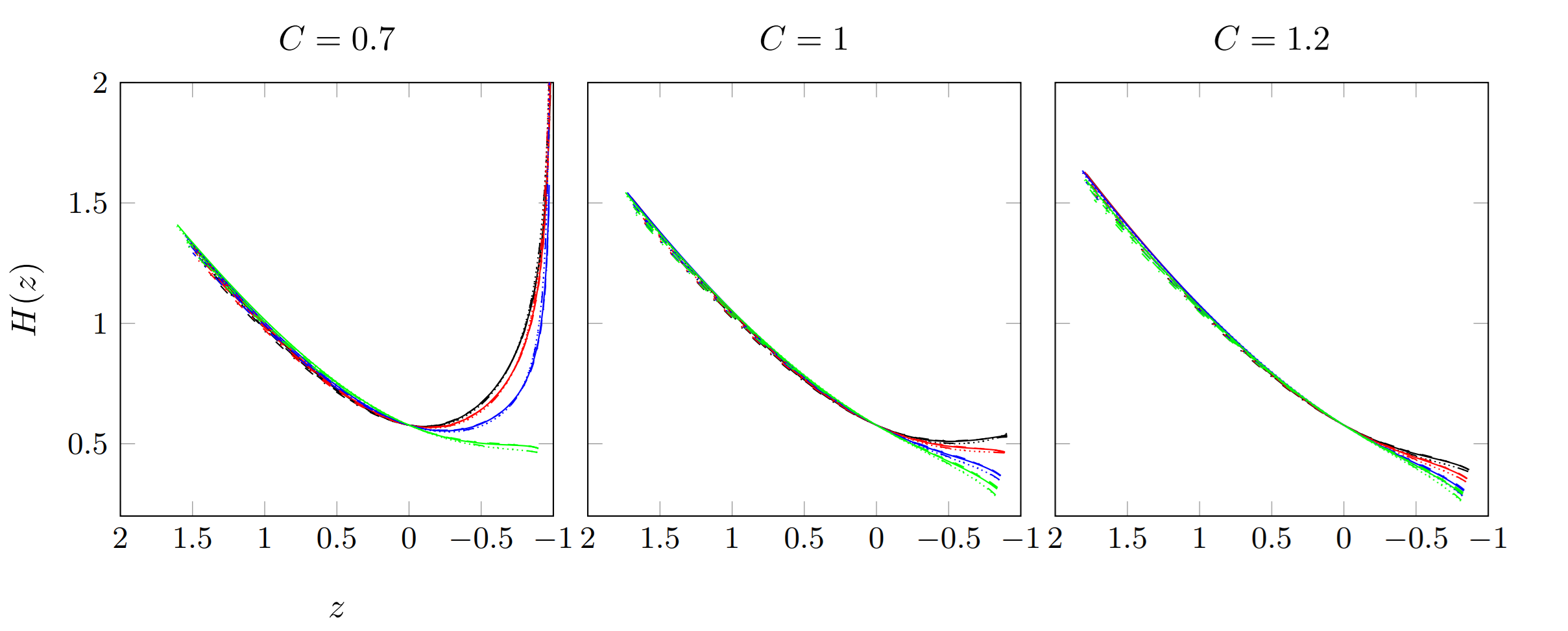}
    \includegraphics[width=1\linewidth]{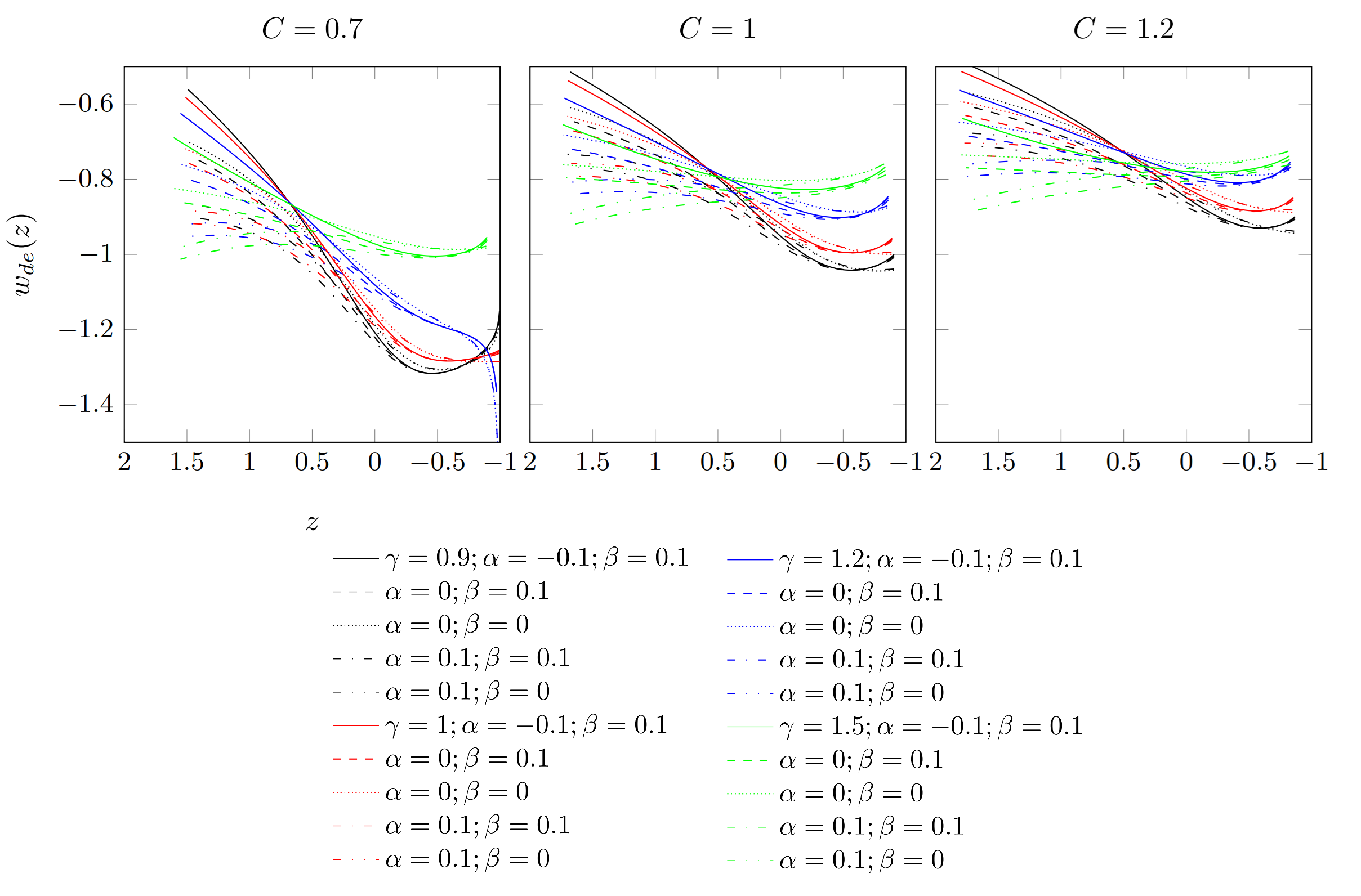}    
    \caption{The dependence of Hubble parameter $H(z)$ (upper panel) and equation-of-state parameter $w_{de}(z)$ (lower panel) in the case of $Q = H(\alpha \rho_m + \beta \rho_{de})$ and for various $C$, $\gamma$ and parameters of interaction $\alpha$ and $\beta$. Although, the dependence of Hubble parameter for $C=0.7$ and $\gamma=0.9$ is similar at first glance to a case without interaction, a final singularity occurs.}
    \label{fig2}
\end{figure}

We see a similar behaviour of Hubble parameter and $w_{de}$ in the case of $C=0.7$ (see fig. 2). Phantomization occurs at some moment in the past for $\gamma=0.9$, 1 and $1.2$. For $\gamma=1.5$ and some values of $\alpha$ and $\beta$, there is a possibility of phantom line crossing in future, and after some time interval, dephantomization.  For $\gamma=1.2$, parameter $w_{de}$ decreases after phantomization, but it decreases more slowly during the initial phase in comparison to $\gamma=0.9$ and $1$. But, in the case of distant future, parameter $w_{de}$ increases for $\gamma=0.9$ and 1, while for $\gamma=1.2$ it sharply decreases. For $\gamma=0.9$, there is a singularity in future. 
For $C=1$ and $\gamma=1$ the phantomization with subsequent dephantomization can take place. The cosmological expansion resembels dynamics without interaction between matter and dark energy. Finally, for $C=1.2$ we do not see significant declinations from the simple model of holographic dark energy. 

\textbf{Model with $Q  = \lambda \frac{\rho_m \rho_{de}}{H}$.} In this case dimensionless function $q$ is
$$
q = 3 \lambda x y
$$
\begin{figure}
    \centering
    \includegraphics[width=1\linewidth]{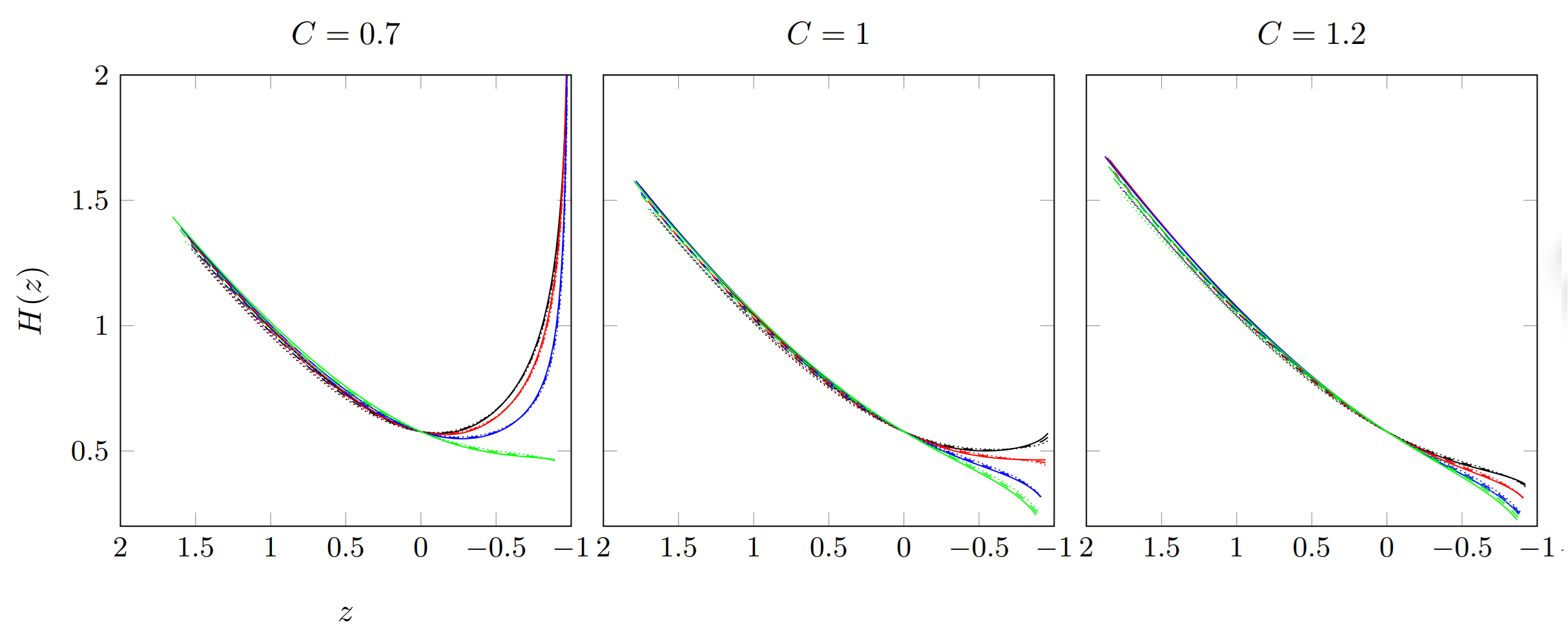}
    \includegraphics[width=1\linewidth]{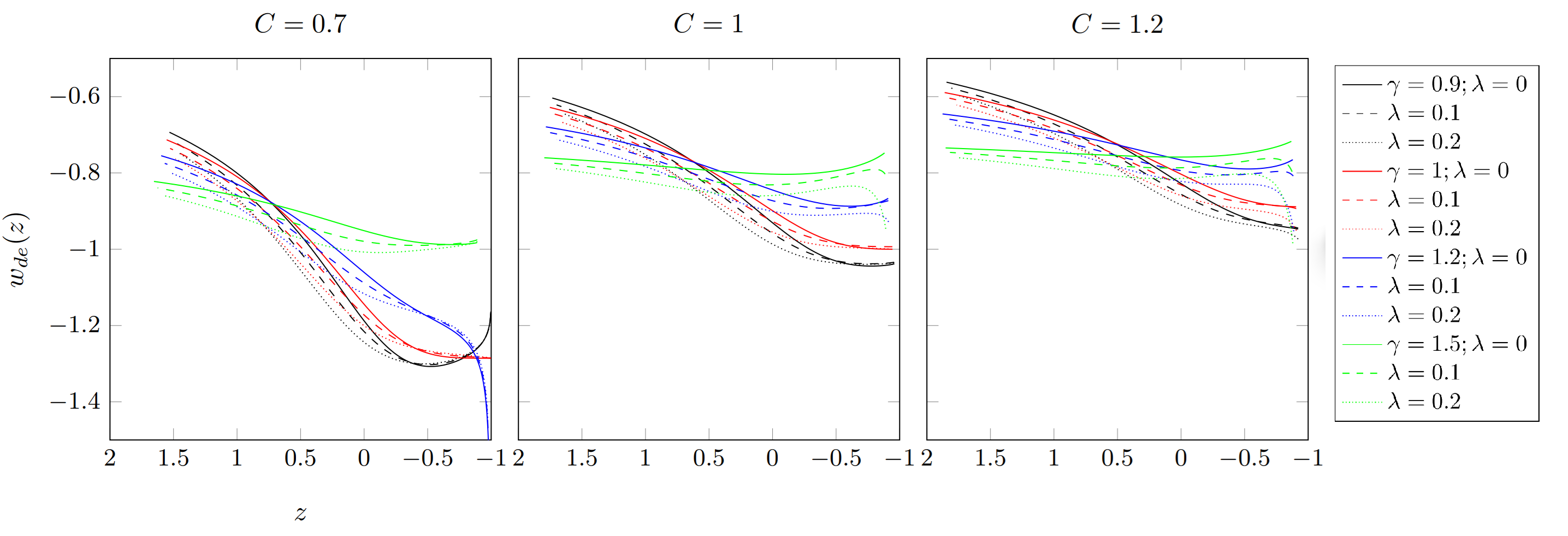}    
    \caption{The same as on fig. 2 for a case of $Q  = \lambda \frac{\rho_m \rho_{de}}{H}$ and for various $C$, $\gamma$ and parameter $\lambda$.}
    \label{fig3}
\end{figure}

In the case of the second interaction (see fig. 3) we note that for large values of $\lambda$ and $C=0.7$, the double crossing of phantom line can appears due to the interaction at $\gamma=1.5$. For $\gamma=0.9$ and $C=0.7$ there is no singularity, and Hubble parameter sharply approaches the constant value.  

\textbf{Model with $Q  = \beta H \rho_{de}^\alpha \rho_{m}^{1 - \alpha}$ ($0 <\alpha < 1$)} is ullustrated on fig. 4. It should be noted that in the case of $\alpha$ from some $\alpha_m$ at fixed $\beta$, the evolution for $\gamma=0.9$ leads to a big rip singularity against asymptotic de Sitter evolution.

\begin{figure}
    \centering
    \includegraphics[width=1\linewidth]{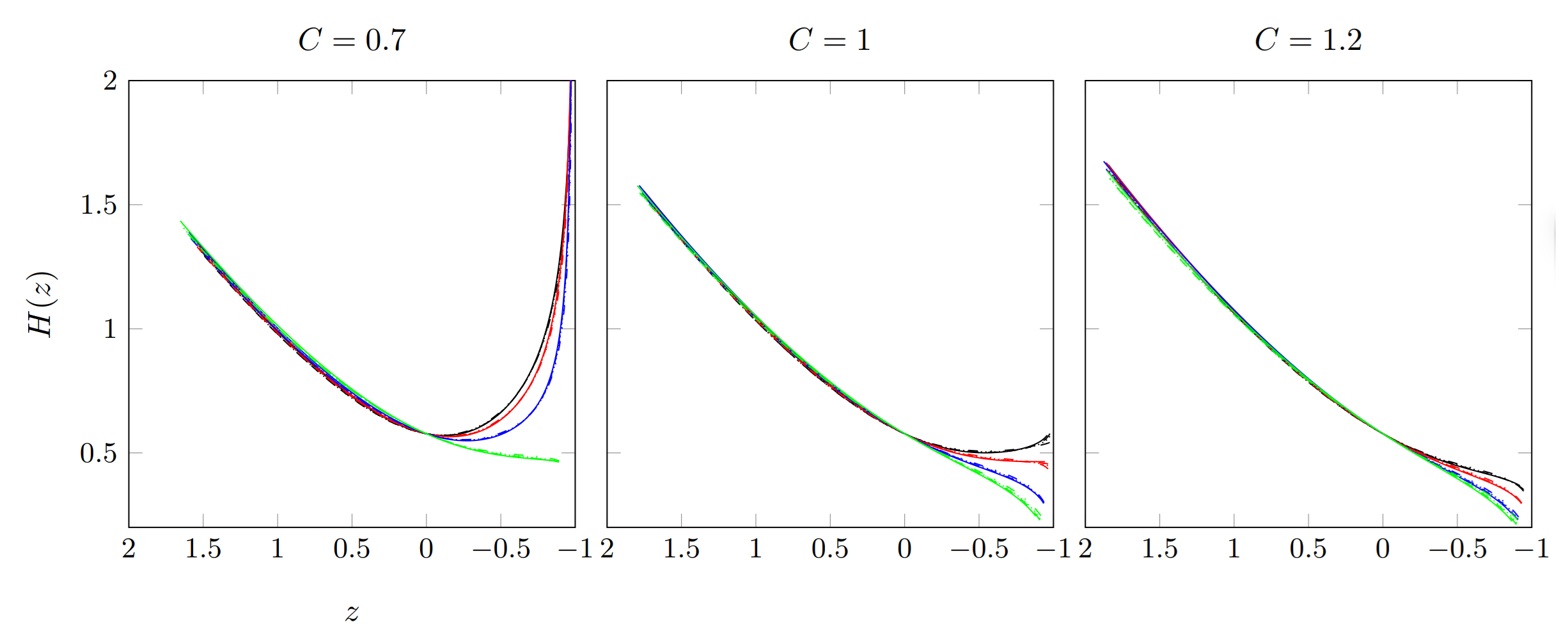}
    \includegraphics[width=1\linewidth]{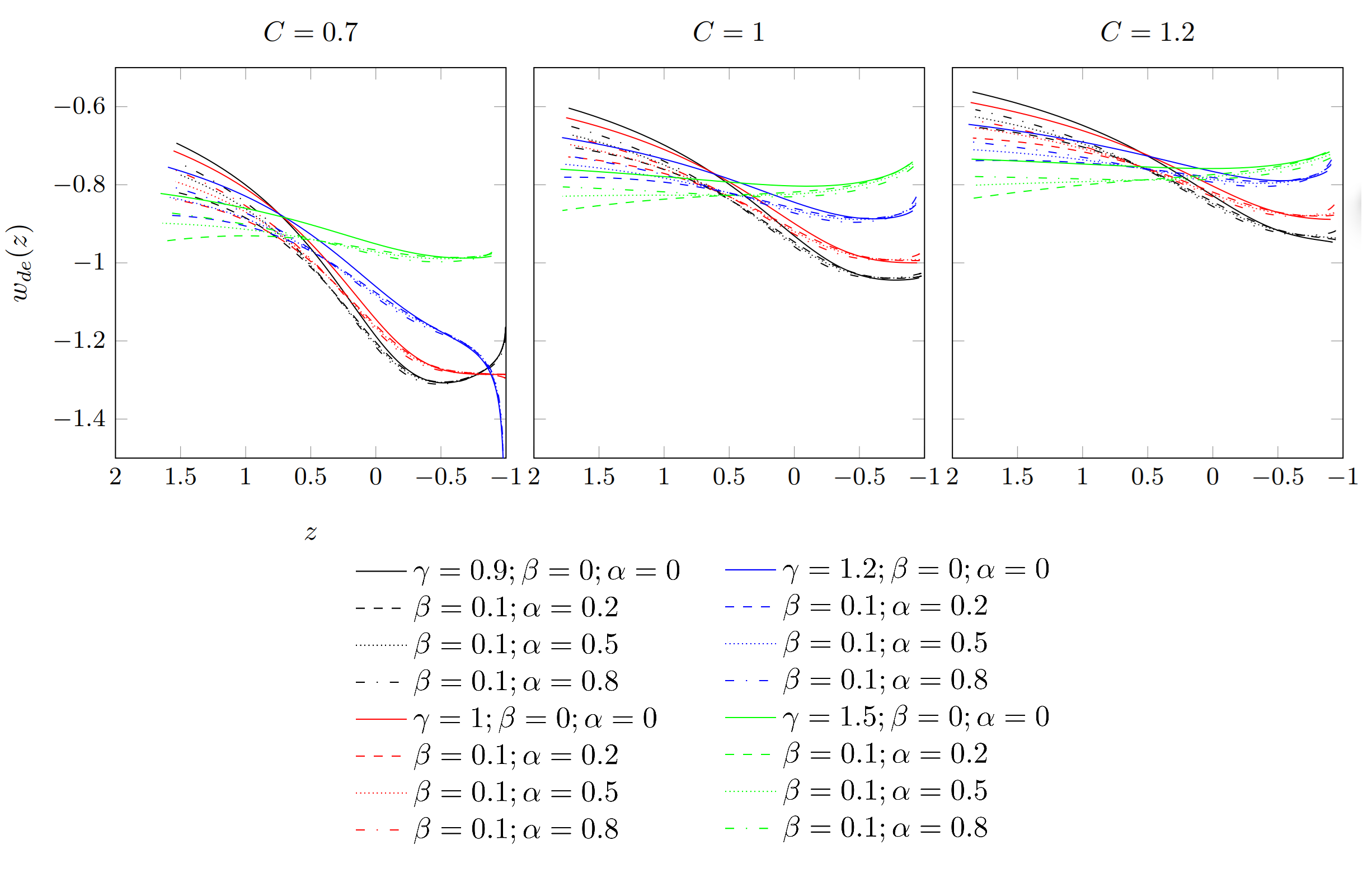}    
    \caption{The same as on fig. 2 for a case of $Q  = \beta H \rho_{de}^\alpha \rho_{m}^{1 - \alpha}$ and for various $C$, $\gamma$ and parameters $\alpha$, $\beta$.}
    \label{fig4}
\end{figure}

\section{Asymptotical evolution and de Sitter attractor/repeller}

It is useful to discuss ``critical points'' of evolution in terms of variables $x$, $y$ and $u$ for holographic dark energy without interaction with matter and for considered interactions.

We consider some early moment of time, when fraction of matter is close to $1$. The expansion of universe leads to decreasing of $x$ and increasing of $y$. A very interesting question is the behaviour of variable $u$.

It is better to start a qualitative analysis with the simplest case, when spacetime is flat and there is no interaction between matter and dark energy. The equation for $u$ allows us to calculate that for $k=0$
\begin{equation}
    \frac{du}{d\eta} = u\left(-\frac{x}{2}+y+(\gamma-2)y(1-u^{-1})\right) - 1.
\end{equation}
It is obvious that for small $y\approx 0$ the derivative is $du/d\eta$<0 for all possible initial values of $u$. Therefore, $u$ starts to decrease. 
The evolution of Universe depends on parameter $\gamma$. One needs to mention that  $\gamma=1$ (standard holographic dark energy model) is peculiar because variable $u$ is defined through the variable $y$. Namely,
$$
y = \frac{C^2}{H^2 L^2} = \frac{C^2}{u^2}
$$
and, therefore, $u_0 = C y_{0}^{-1/2}$. The asymptotical value for $u$ is $C$ for large times. We know that $C<0$ corresponds to a big rip singularity in the future. The evolution of scale factor in the case of such singularity is
\begin{equation}
    a = \frac{a_c}{(t_f - t)^\epsilon}, \quad \epsilon>0.
\end{equation}
Simple calculations show that $u$ for this scenario is constant and
$$
u_{\infty} = \frac{\epsilon}{\epsilon + 1}<1.
$$
The values of $C>1$ otherwise correspond to asymptotical expansion according to the law $a\sim t^{\epsilon}$, $\epsilon>1$. For this case, $u$ approaches constant value
$$
u_{\infty} = \frac{\epsilon}{\epsilon - 1}>1.
$$
Expansion, according to de Sitter law for $t\rightarrow\infty$, corresponds to $u_{\infty} = 1$.

If $\gamma\neq 1$ relation between $u$ and $y$ depends on Hubble parameter $H$, then
\begin{equation}
    u = \left(\frac{C^2}{y}\right)^{\frac{1}{4-2\gamma}}H^{\frac{1-\gamma}{2-\gamma}}
\end{equation}
or
$$
u = \frac{C}{y^{\frac{1}{4-2\gamma}}}\left(\frac{H}{C}\right)^{\frac{1-\gamma}{2-\gamma}}.
$$
Therefore, we can consider various initial conditions for $u$. We can define
$$
\gamma = 1 + \chi, 
$$
where $|\chi|$<<1. For times when $y\approx 1$, we have that
$$
\frac{du}{d\eta} \approx u\chi \left(1-\frac{1}{u}\right).
$$
One can conclude that for $\gamma<1$ ($\chi<0$) there is attractor $u = 1$ i.e. evolution of universe asymptotically approaches de Sitter regime.

The situation changes for $\gamma>1$. Therefore, we can expect two possible cases for time dependence of $u$. Initially, ($x\approx 1$) $u$ decreases. For sufficiently large $u_{0}$, decreasing can change to increasing for $u>1$. If $u(\eta)$ intersects line $u=1$, then $u$ decreases to 0.  From relation for $u$ it is evident that decreasing $u$ corresponds to increasing of Hubble parameter i.e. universe ends its existence in a  singularity. On the other hand, increasing of $u$ for $\gamma>1$ indicates that $H\rightarrow 0$ i.e. the rate of universe expansion decreases.  

\begin{figure}
    \centering
    \includegraphics[width=0.9\linewidth]{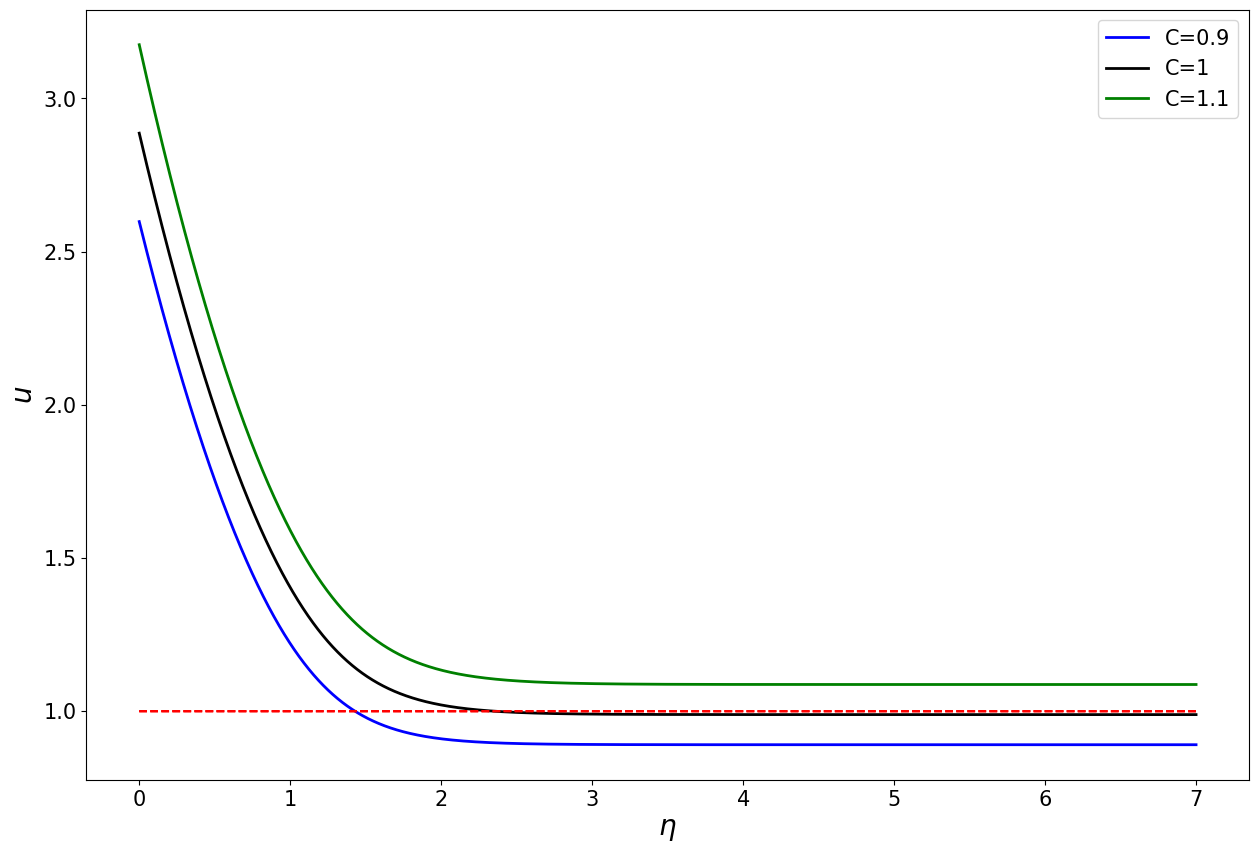}
    \caption{The dependence of $u$ from dimensionless time $\eta$ for $\gamma=1$ without interacton between matter and holographic dark energy. For $\eta=0$, we put $x=0.9$, $\Omega_k = -0.02$. The line $u=1$ corresponding to de Sitter solutios is also depicted. }
    \label{0}
\end{figure}

\begin{figure}
    \centering
    \includegraphics[width=0.9\linewidth]{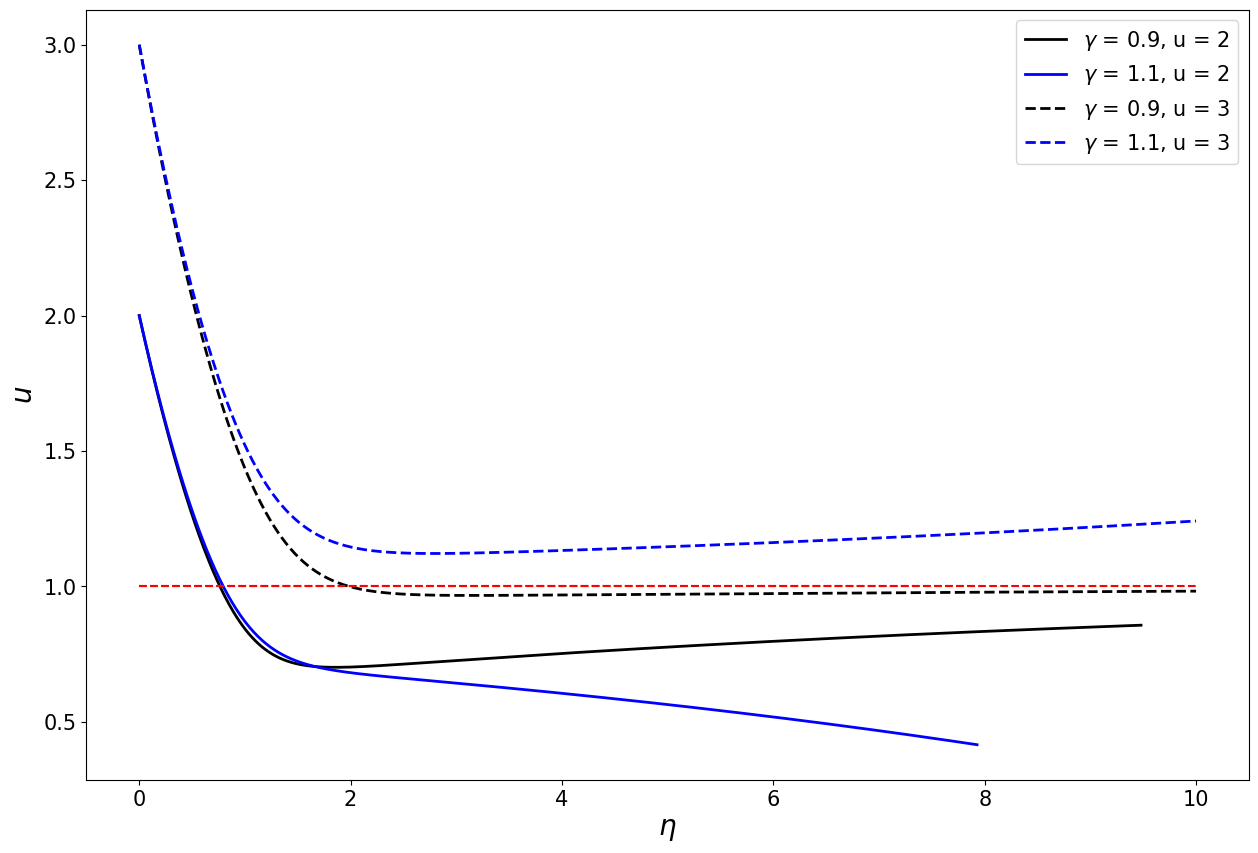} 
    \caption{The same as on fig. 5 but for $\gamma\neq 1$. For $\gamma<1$ the point $u=1$ in distant future is an attractor: universe approaches de Sitter regime for $\eta\rightarrow\infty$. But for $\gamma<1$ $u=1$ is the saddle point: there are two possible trajectories for $u$. One of them ($u$ approaches 0 for finite conformal time) corresponds to singularity in future. Increasing of $u$ after some minimum $u_{min}>1$  corresponds to expansion with $H\rightarrow 0$ for $\eta\rightarrow \infty$. }
    \label{0-1}
\end{figure}

Consideration of model with nonzero curvature does not change these issues because, for large times, $\Omega_k$ is very close to $0$ and therefore we can apply our analysis below. On figs. 5 and 6 we depicted evolution of function $u(\eta)$ for $\gamma=1$ and $\gamma\neq 1$ correspondingly. We start from the values $x=0.9$ and $\Omega_k =  0.02$ at $\eta=0$ as an example.

The next step is to consider the case of interaction between matter and dark energy. For interaction of the first type, where $Q=H(\alpha \rho_m + \beta\rho_{de})$, our results are given on fig. 7 for $\beta>0$. For $\gamma=1$,  due to the interaction, universe ends its existence in singularity for all values of $C$. The function $u$ monotonically decreases. If $\gamma<1$, there is an attractor $u_{\infty}<1$ and, therefore, the big rip singularity occurs. There is no asymptotic de Sitter expansion for $\gamma<1$. For $\gamma>1$ there are two possible scenarios. One of them corresponds to a singularity more rigid in comparison to the big rip and the second is evolution with $H\rightarrow 0$ at $\eta\rightarrow \infty$. 

\begin{figure}
    \centering
    \includegraphics[width=0.7\linewidth]{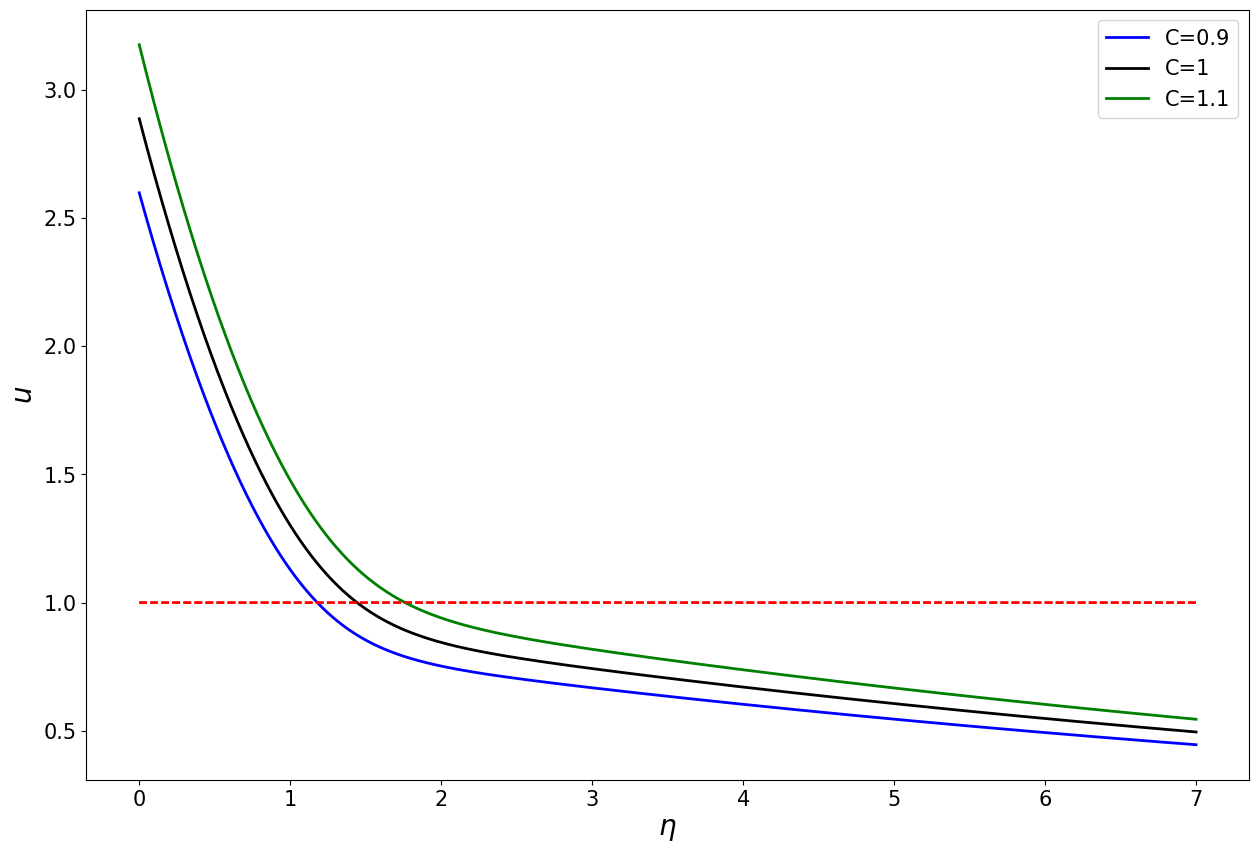} 
    \includegraphics[width=0.7\linewidth]{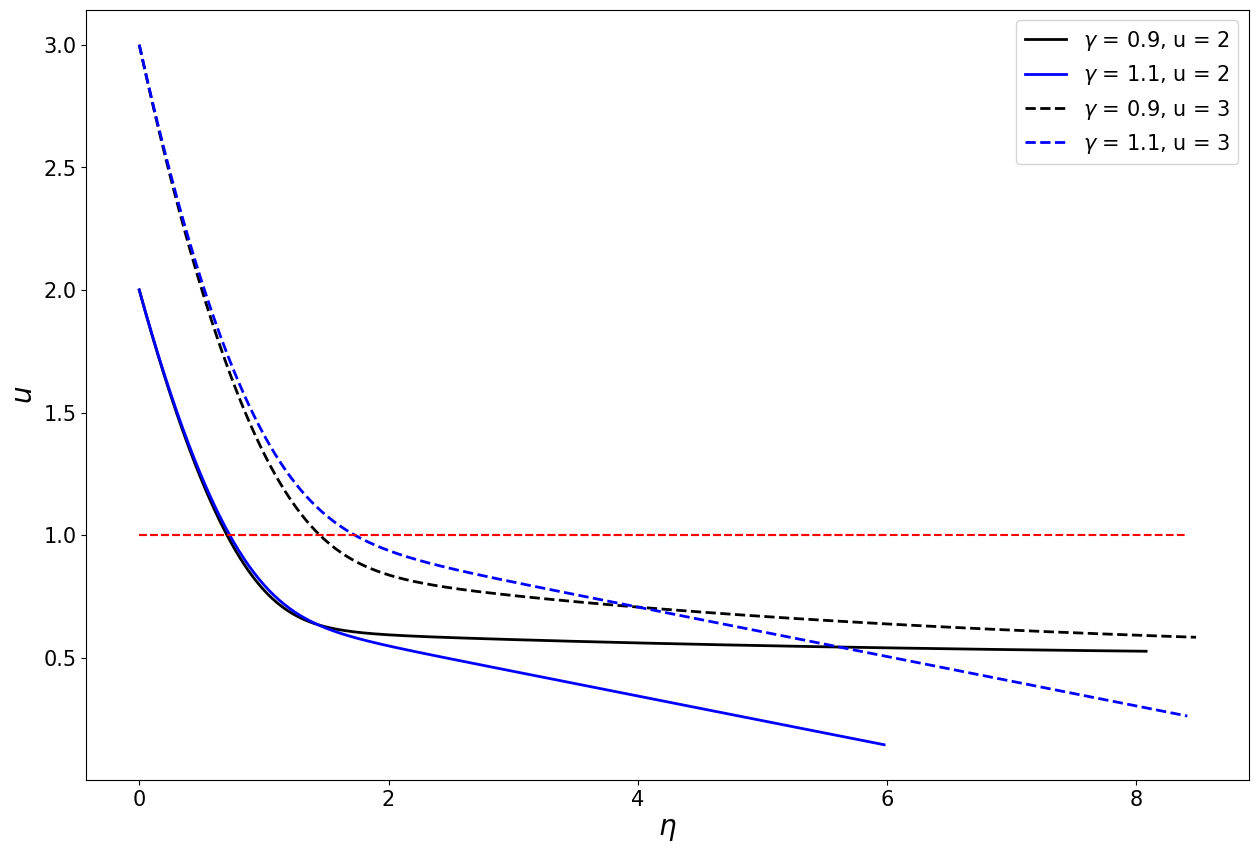}
    \caption{The evolution of function $u$ for interaction with $Q=H(\alpha \rho_m + \beta\rho_{de})$ ($\alpha=\beta=0.1$) in a case of $\gamma=1$ (upper panel) and $\gamma\neq 1$ (lower panel).}
    \label{1}
\end{figure}

\begin{figure}
    \centering
    \includegraphics[width=0.7\linewidth]{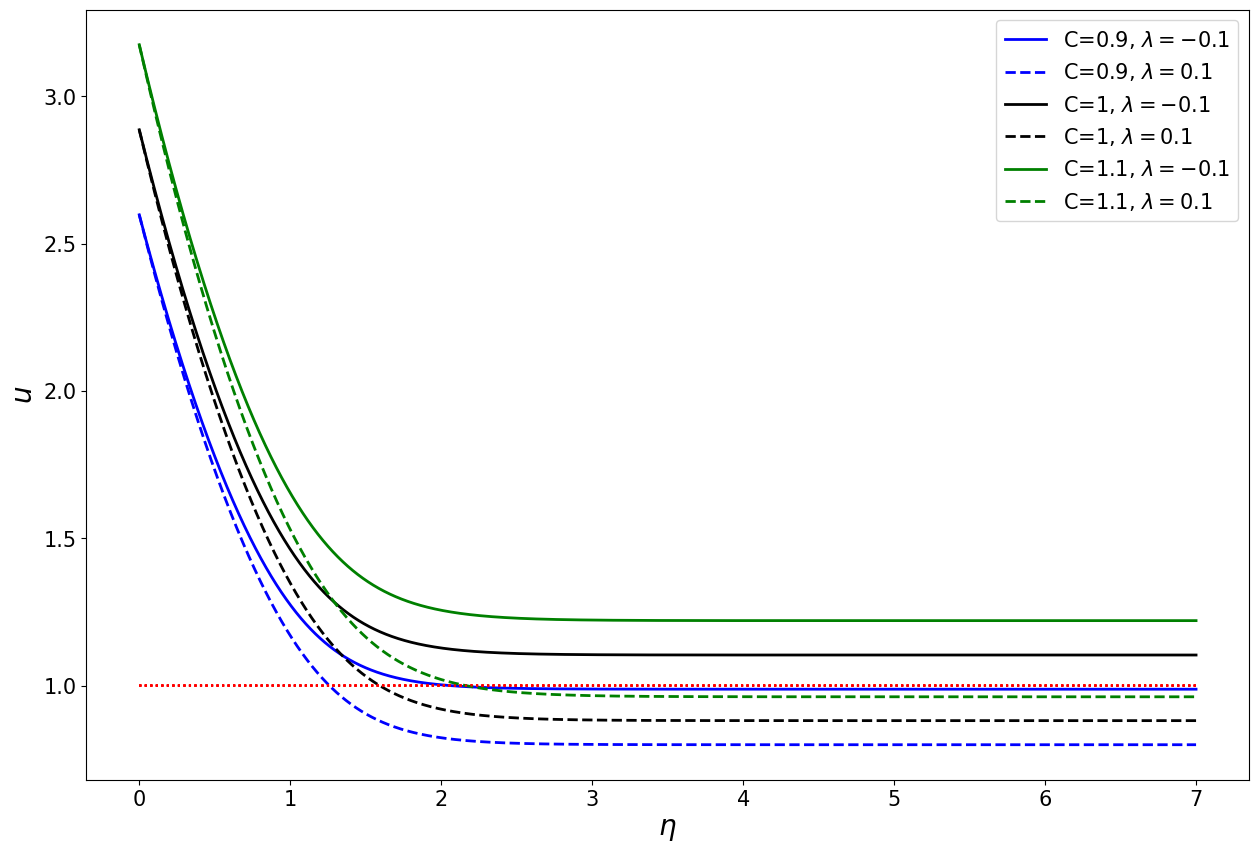} 
    \caption{The evolution of $u(\eta)$ for $\gamma=1$ and for interaction with $Q = \lambda \rho_m \rho_{de}/H$. Negative values of $\lambda$ soften or cancel future singularity.}
    \label{2-2}
\end{figure}

\begin{figure}
    \centering
    \includegraphics[width=0.7\linewidth]{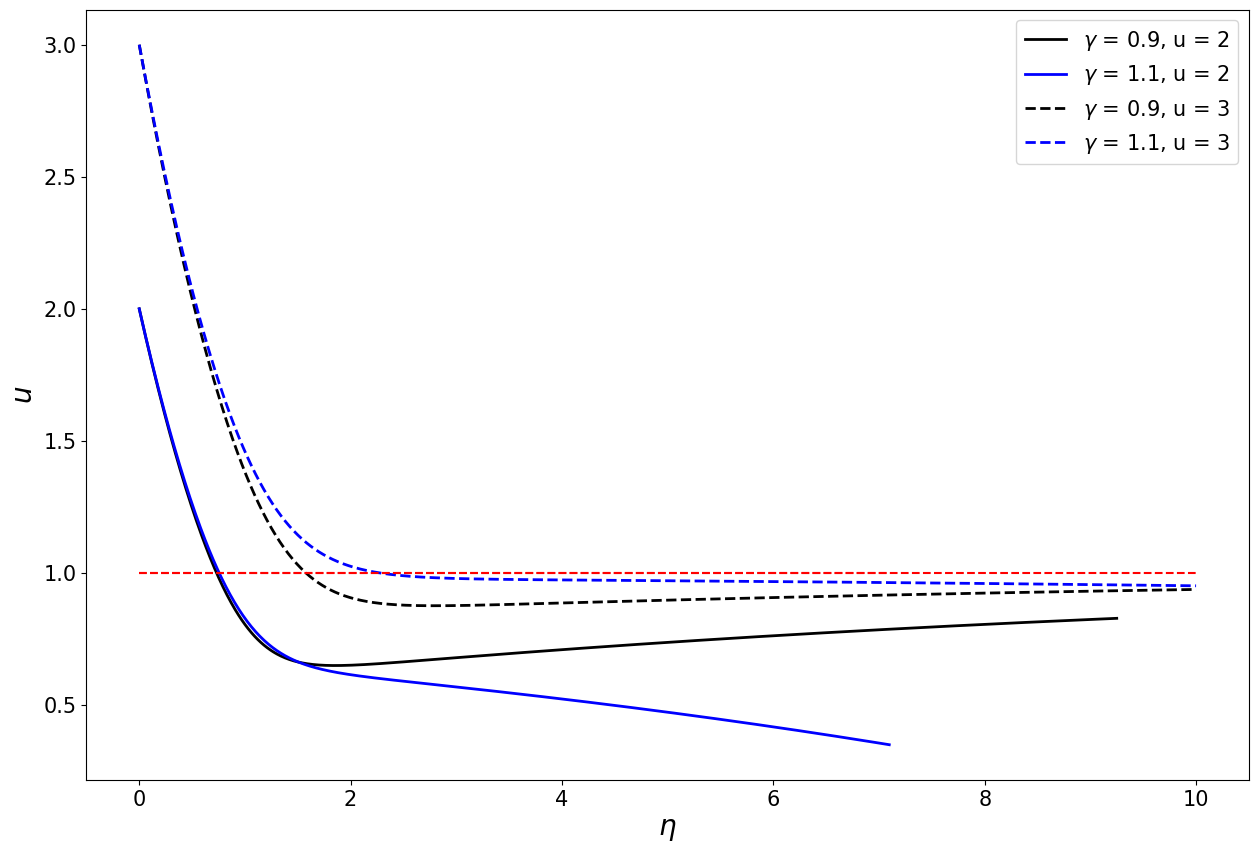} 
    \includegraphics[width=0.7\linewidth]{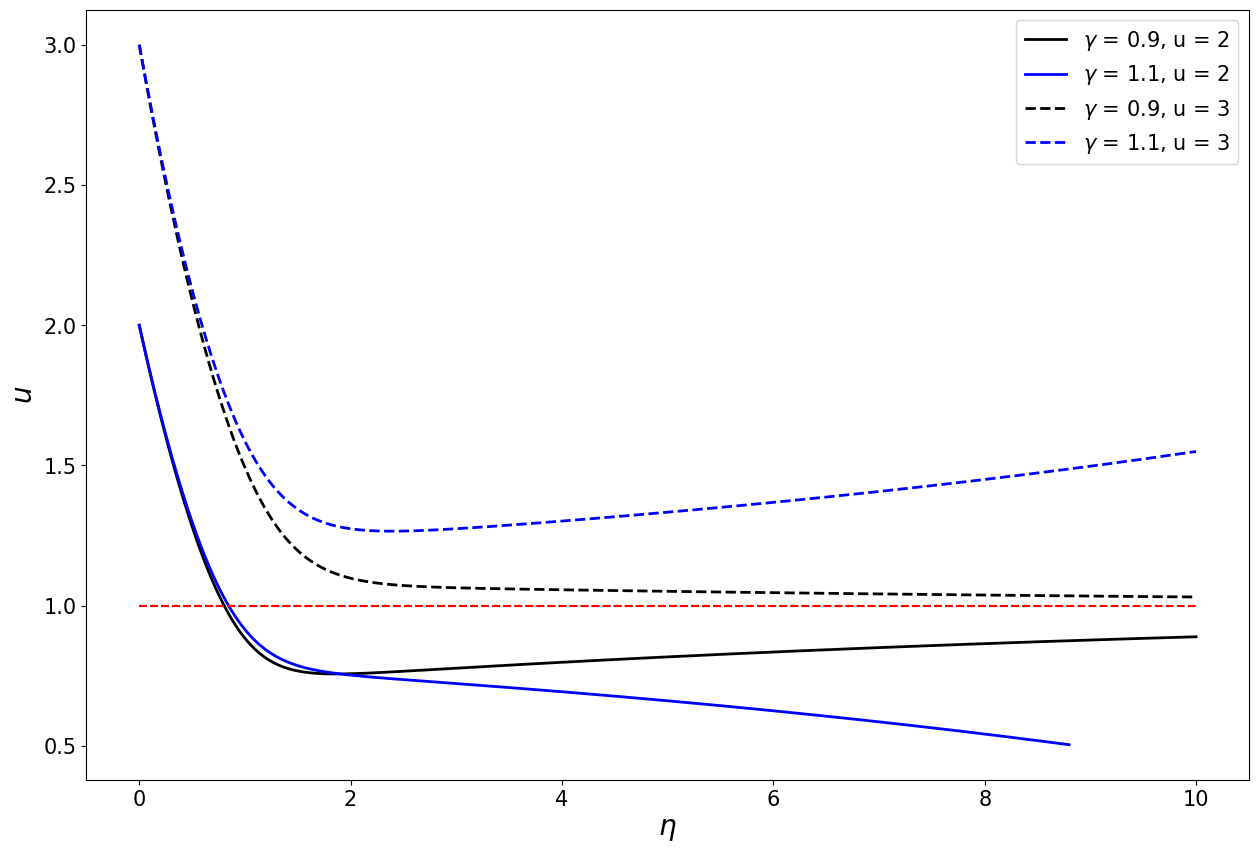}
    \caption{The same as on fig. 8 but for $\gamma\neq 1$: $\lambda=0.1$ (upper panel) and $\lambda=-0.1$ (lower panel). For $\lambda=-0.1$, there is no singularity in the future for $\gamma=1.1$ and $u_0 = 3$, which take place at $\lambda=0.1$.}
    \label{2}
\end{figure}

The next variant of interaction with $Q = \lambda \rho_m \rho_{de} / H$ leads to a possibility of elimination of singularities for negative values of $\lambda$ at $\gamma=1$ (see fig. 8). Also, asymptotic de Sitter expansion can change to expansion according to the power law with $H\rightarrow 0$. Otherwise, values $\lambda>0$ lead to a big rip singularity in the future for all $C$. On Fig. 9, the case of $\gamma\neq 1$ is presented. For $\gamma<1$, the point $u=1$ is an attractor similarly to the case without interaction. For $\gamma>1$, interaction at negative $\lambda$ can eliminate the singularity occuring at $\lambda\geq 0$. 

\begin{figure}
    \centering
    \includegraphics[width=0.7\linewidth]{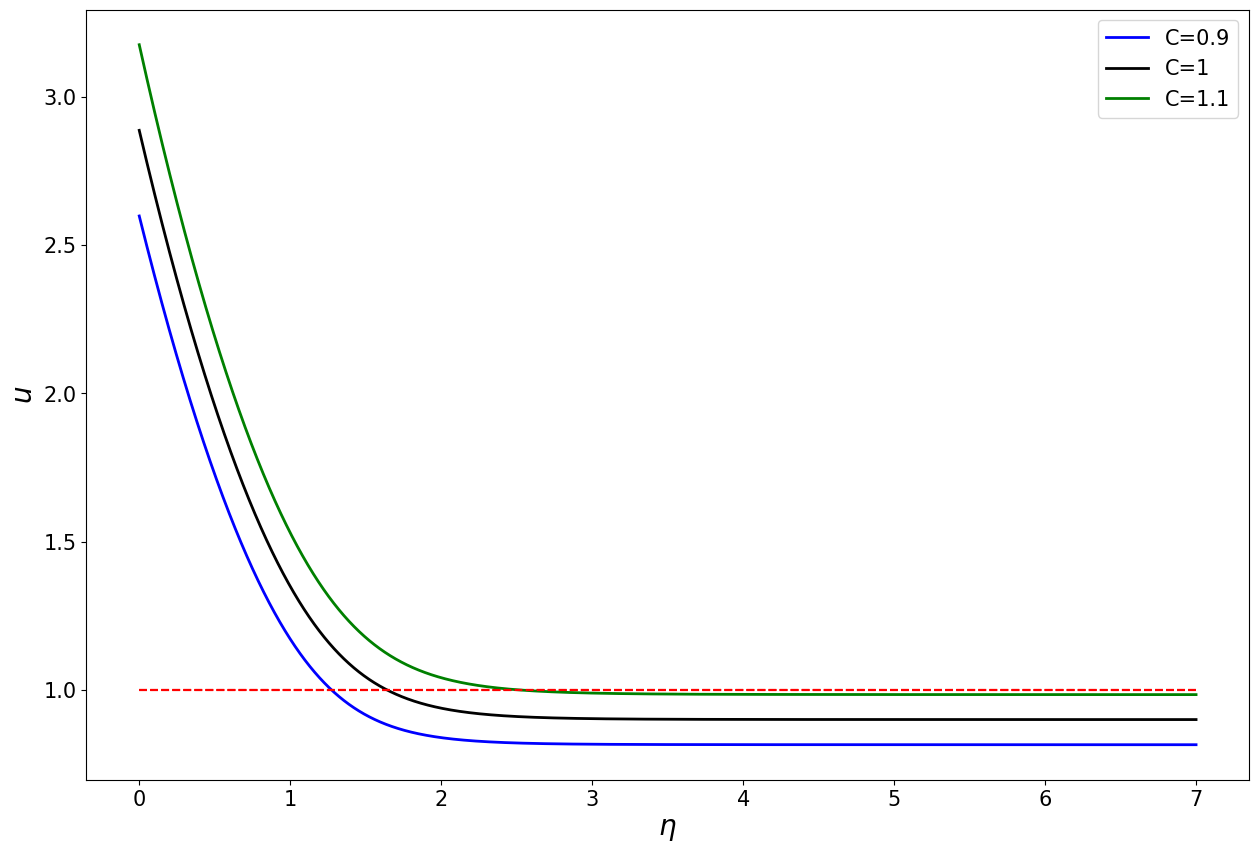} 
    \includegraphics[width=0.7\linewidth]{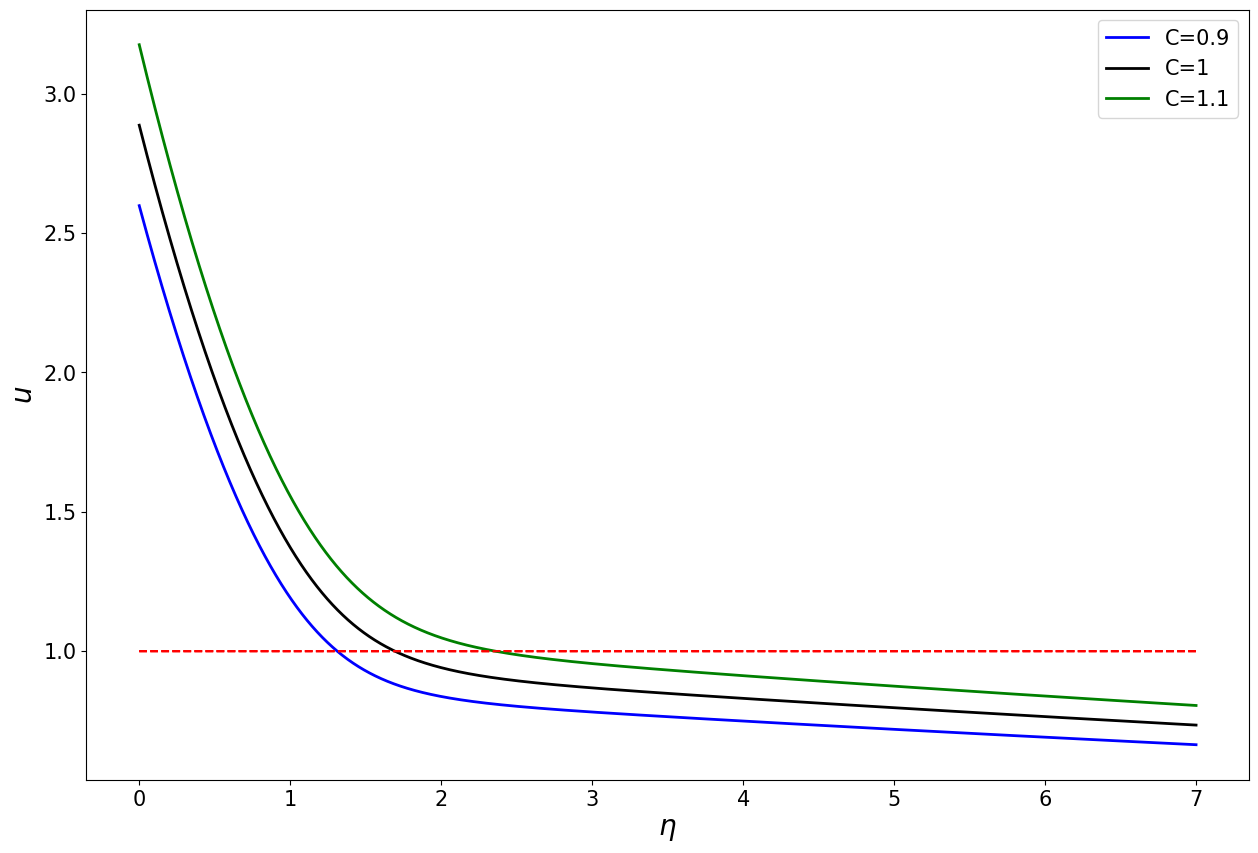}
    \caption{The same as on fig. 8 for  $Q = \beta H \rho_{m}^{\alpha}\rho_{de}^{1-\alpha}$ in the case of $\beta=0.1$, $\alpha=0.3$ (upper panel) and $\beta=0.1$, $\alpha=0.8$ (lower panel). The singularity can appear for $C>1$. For $\alpha=0.8$ universe expands faster in comparison with big rip regime for simple phantom model. }
    \label{3}
\end{figure}

\begin{figure}
    \centering
    \includegraphics[width=0.7\linewidth]{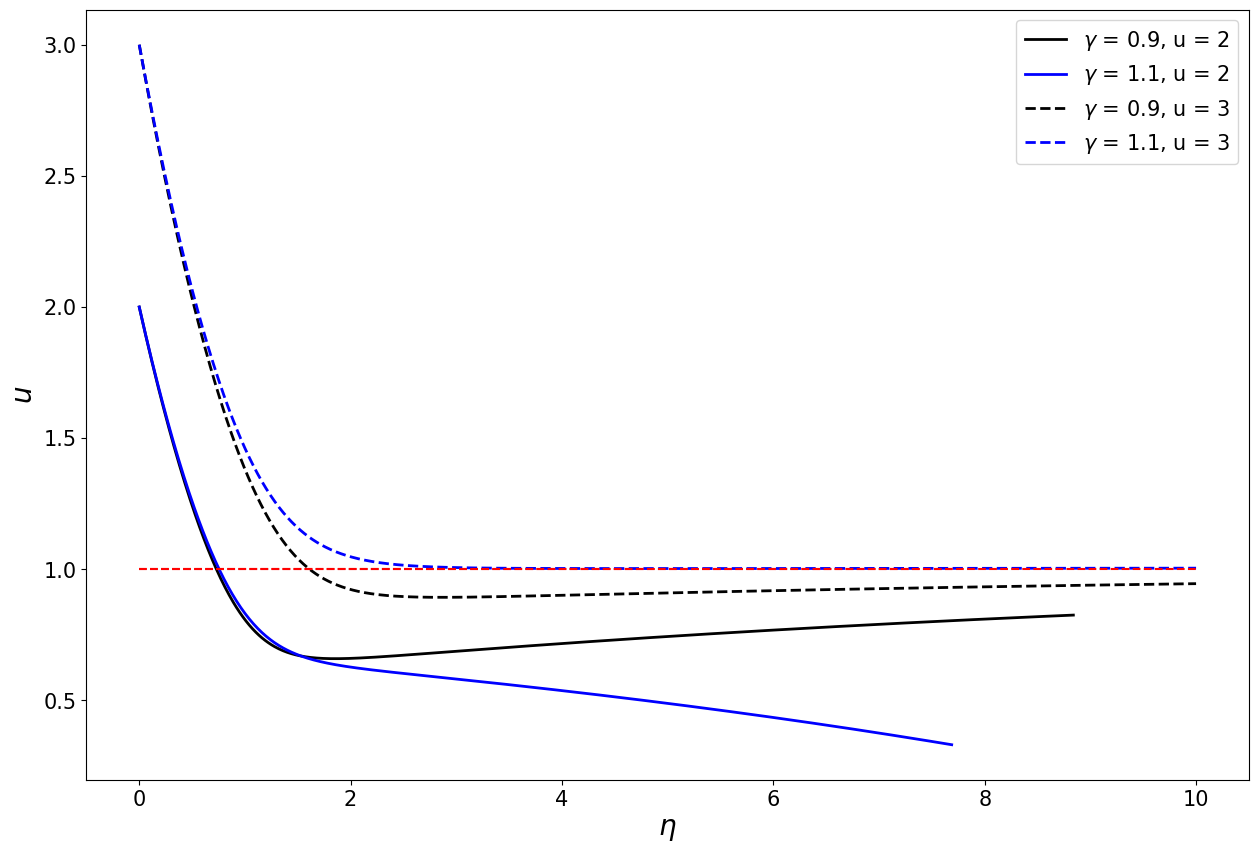} 
    \includegraphics[width=0.7\linewidth]{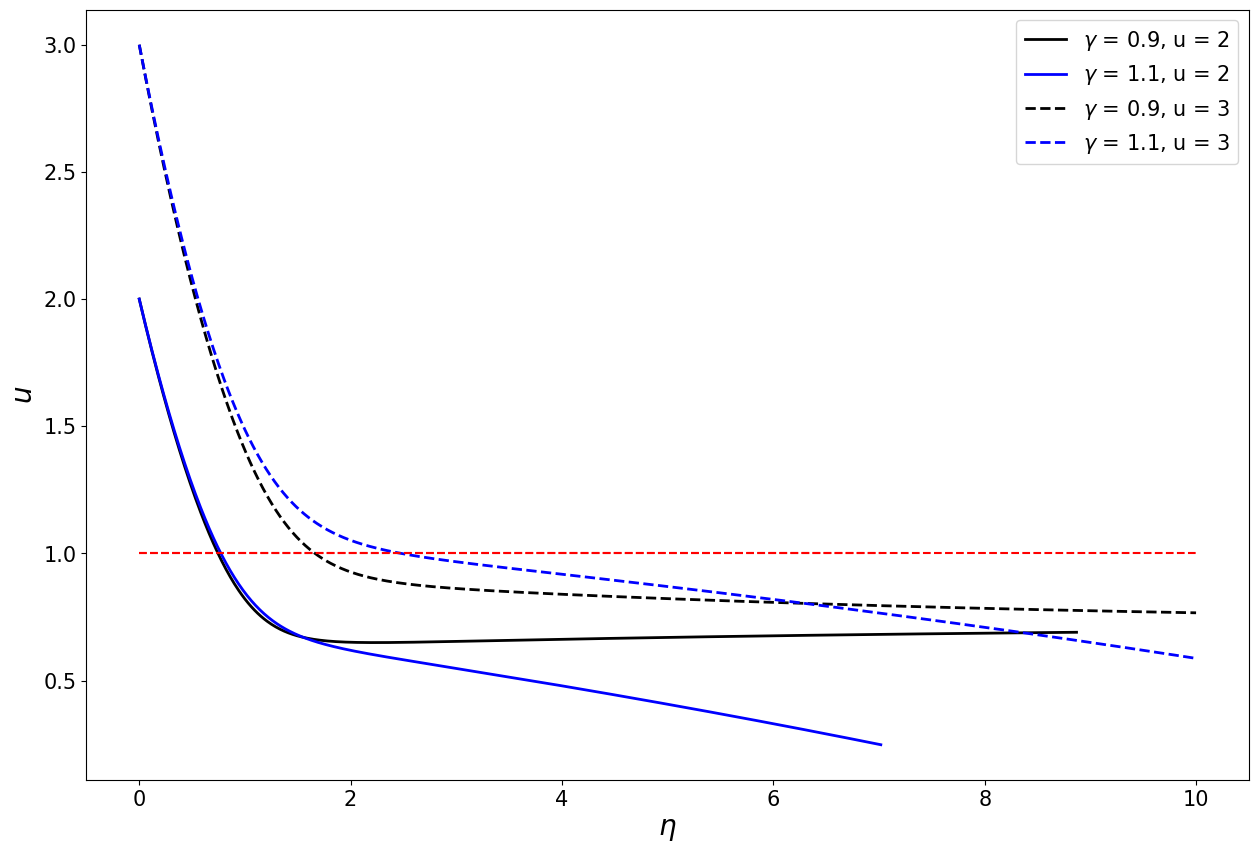}
    \caption{The evolution of $u(\eta)$ for $\gamma\neq 1$ and $Q = \beta H \rho_{de}^{\alpha}\rho_{m}^{1-\alpha}$ in the case of $\beta=0.1$, $\alpha=0.3$ (upper panel) and $\beta=0.1$, $\alpha=0.8$ (lower panel). For the first case at $\gamma=0.9$, the asymptotic de Sitter expansion occurs, but for the second case a big rip singularity takes place.}
    \label{4}
\end{figure}

Finally, we investigated the interaction with $Q = \beta \rho_{de}^{\alpha}\rho_m^{-\alpha} $ ($\alpha>0$). Again one needs to discriminate between the cases $\gamma=1$ and $\gamma\neq 1$. We consider $\beta>0$ and see that the interaction leads to singularities for various $C$, including $C>1$, and $\gamma=1$ (fig. 10). From some value of $\alpha$, singularity occurs for any $C$. For $\gamma\neq 1$ (fig. 11), the existence of attractor $u=1$ depends on $\beta$ and $\alpha$. For fixed $\beta$, the non-singular expansion due to the interaction turns into singular for some $\alpha$.

\section{Conclusion}

We investigated the possible scenarios of evolution for the universe filled by Tsallis hologaphic dark energy interacting with matter. Without interacting, for $\gamma<1$, we have asymptotic de Sitter expansion, although, Hubble parameter may grow sufficiently rapidly as in the case of $\gamma=1$ and $C<1$, when a big rip singularity occurs. For $\gamma>1$, de Sitter solution appears as the repeller. For some initial conditions, the universe expands so that it ends its existence in a singularity. Another solution corresponds to expansion with $H\rightarrow 0$ at $t\rightarrow \infty$. 

Interaction between matter and holographic dark energy changes these issues. We considered three possibilities for function $Q$ ($-Q$), appearing in r.h.s. of continuity equation for holographic dark energy (matter). For the simple case $Q = H(\alpha\rho_m + \beta\rho_{de})$ with $\beta>0$, the singularity of a big rip takes place for all $C$ at $\gamma=1$. The de Sitter attractor solution for $\gamma<1$ also disappears and for $\gamma<1$ a big rip also occurs. For $\gamma>1$ the non-singular solution is possible. For the second type of interaction with $Q = \lambda\rho_m \rho_{de}/H$ at $\lambda<0$ a future singularity in the case of $\gamma=1$ can be softened or cancelled. For $\gamma<1$, once again, there is an asympotical de Sitter solution similar to the non-interacting case. Finally for $Q = \beta\rho_{de}^{\alpha}\rho_{m}^{1-\alpha}$ we can see another features. The de Sitter attractor can dissapear for $\gamma<1$ and for some $\alpha$ at fixed value of $\beta$.

Therefore the simple model of holographic dark energy with non-additivity parameter $\gamma\neq 1$ can demonstrate a sufficiently rich repertoire of behavior. The existence of asymptotical de Sitter regime for $\gamma<1$ (in a case without interaction between matter and dark energy and for some interactions) can witnesses in favor of assumption that holographic dark energy is a real alternative to $\Lambda$CDM model.

\section*{Acknowledgments}

This research was supported by funds provided through the Russian Federal Academic Leadership Program “Priority 2030” at the Immanuel Kant Baltic Federal University

\end{document}